\documentclass[aps,twoside,twocolumn,showpacs,showkeys]{revtex4}
\usepackage{graphicx}
\usepackage{amsmath}
\begin{document}
\title[]{General Solution of EM Wave Propagation in Anisotropic Media}
\author{Jinyoung \surname{Lee}}
\email{arameas@kaist.ac.kr}
\affiliation{Electrical and Electronic Engineering Department, Korea Advanced Institute of Science and Technology, Dajeon 305-714}
\author{Seoktae \surname{Lee}}
\email{stlee@semyung.ac.kr}
\affiliation{Electronic Engineering Department, Semyung University, Jecheon 390-711}
\date[]{}


\def\a{\alpha}               \def\b{\beta}
\def\d{\delta}               \def\D{\Delta}
\def\e{\epsilon}
\def\f{\phi}                 \def\F{\Phi}
\def\g{\gamma}            \def\G{\Gamma}
\def\j{\psi}                  \def\J{\Psi}
\def\l{\lambda}             \def\L{\Lambda}
\def\m{\mu}                 \def\n{\nu}
\def\o{\omega}               \def\O{\Omega}
\def\p{\pi}                  \def\P{\Pi}
\def\r{\rho}
\def\s{\sigma}              \def\S{\Sigma}
\def\t{\tau}
\def\th{\theta}             \def\Th{\Theta}

\def\etal{{\it et al.}}

\def\sl #1{\not\!{#1}}

\def\abs#1{\left| #1\right|}
\def\Bar{\overline}

\def\fr{\frac}
\def\cd{\cdot}
\def\ba{\begin{array}}          \def\ea{\end{array}}
\def\bz{\begin{equation}}       \def\ez{\end{equation}}
\def\by{\begin{eqnarray}}       \def\ey{\end{eqnarray}}
\def\pa{\partial}               \def\na{\nabla}
\def\tb{\textbf}
\def\nn{\nonumber}              \def\ni{\noindent}

\makeatletter
\newcommand{\rmnum}[1]{\romannumeral #1}
\newcommand{\Rmnum}[1]{\expandafter \@slowromancap\romannumeral #1@}
\makeatother

\begin{abstract}
When anisotropy is involved, the wave equation becomes simultaneous
partial differential equations that are not easily solved.
Moreover, when the anisotropy occurs due to both permittivity and permeability, these equations are insolvable without a numerical or an
approximate method. The problem is essentially due to the fact neither
$\e$ nor $\m$ can be extracted from the curl term, when they are in it. The
terms $\na\times {\bf E}$ (or ${\bf H})$ and $\na\times \e{\bf
E}$ (or $\;\m{\bf H})$ are practically independent variables, and $\bf
E$ and $\bf H$ are coupled to each other. However, if Maxwell's equations
are manipulated in a different way, new wave equations are
obtained. The obtained equations can be applied in anisotropic, as well as
isotropic, cases. In addition, $\bf E$ and $\bf H$ are decoupled in
the new equations, so the equations can be solved analytically by using tensor Green's functions.
\end{abstract}

\maketitle
\section{Introduction}
As the importance of anisotropic devices has increased in many fields of
optics and microwaves, wave propagation in anisotropic media has been
widely studied over the last decades \cite{1}. The anisotropic nature
basically stems from the polarization or magnetization that can occur in
materials when external fields pass by. Generally, certain axis
components of the $\bf E$ and the $\bf H$ fields are influenced by other axis components
and by those of the same axis. This is why matrices are involved in $\e$ and $\m$.  Therefore, Maxwell's equations and the wave equations are also represented in matrix form. Mathematically the electric field $\bf E$ and the magnetic field $\bf H$ are not only vectors, but also rank-1 tensors, which implies that they obey the set of rules of coordinates transformation. This also implies that $\e$  and $\m$ are tensors of rank-2.\\

In ordinary homogeneous isotropic media, a wave equation with a source term is solved with an ordinary adaptation of Green's functions. However, in anisotropic media, the equations become linear simultaneous partial differential equations (PDE). These equations contain all information pertaining to the anisotropic properties. As they are linear, Green's functions can be considered easily with the source terms. These types of solutions with Green's functions  have been studied under the name of "dyadic Green's functions" over the last few decades. Most of published papers presented the special properties of applications or methods of obtaining Green's functions in a specified set of coordinates.\\

This paper presents a general methodology of solving the
wave propagation equation in an anisotropic environment
to obtain the \tb{E} and the \tb{H} fields. The tensor Green's functions $GE_{ij}$  and
$GH_{ij}$ for an electric and a magnetic field are used, as these
equations are linear. Generally for an unbounded case, a Fourier-transformed Green's function is useful. It changes differential
equations into algebraic equations in $\bf{k}$-space. Therefore, finding
the Fourier-transformed Green's functions is the core procedure in 
solving the wave equations.\\

In Section \Rmnum{2}, the anisotropic characteristics are only concerned with the permittivity $\e$, which becomes a matrix. The corresponding Fourier-transformed Green's functions for this case are easily obtained. Therefore the equation
system is solvable. The Section \Rmnum{3} addresses a case in which anisotropy
is described only with the matrix permeability $\m$. The mathematical procedure in this case is similar to that discussed in Section \Rmnum{2}. The subsequent section, Section \Rmnum{4}, is an arbitrary case. The anisotropy comes from both the permittivity and the permeability. The equations appear to be insolvable in an analytical sense. Other algebraic manipulations of Maxwell's equations lead to new wave equations, in which $\bf E$ and $\bf H$ are decoupled. Analytic solutions are also possible with the use of Green's functions. The last section contains the discussion and conclusion.

\section{ANISOTROPY FROM PERMITTIVITY}
The typical method for deriving of the wave equation starts from

\by \na \times \na \times {\bf E} &=&  - \frac{\pa
}{{\pa t}}(\na  \times \mu {\bf H}), \\
\na \times \na \times {\bf H} &=& \na \times \bf J +
\frac{\pa}{\pa t} (\na \times \e \bf E). \ey

\ni In an ordinary homogeneous isotropic case, the constants $\e$ and $\m$ are extracted from the curl term, and the equations lead to wave equations. In this section, it is assumed that anisotropy exists due to the permittivity. Additionally, $\e$ is a matrix, but $\m$ is a scalar number. Thus, Eq.(1) is considered as an isotropic case. On the other hand, Eq.(2) has a problem in the curl term on the right side. The matrix $\e$ cannot come out of the curl term, and Maxwell's 3rd equation cannot be applied. Therefore, the equations become

\by &&\na  \times \na  \times {\bf E} + \frac{1}{c^2 }\mu _r
\e_r \frac{\pa ^2 {\bf E}}{\pa t^2 } =  -
\mu _0 \mu _r \frac{\pa {\bf J}}{\pa t},\\
&&\na ^2 {\bf H} =  - \na  \times {\bf J} - \e_0
\frac{\pa }{\pa t}(\na  \times \epsilon _r {\bf E}). \ey

\ni Here the first term of Eq.(3) was not expanded to $\na (\na  \cdot {\bf
E}) - \na ^2 {\bf E}$, as the divergence of $\bf E$ cannot be
written as $\e^{-1} \r$ if using Maxwell's equation. When $\e$ is a matrix, the correct form of the first Maxwell's equation is $\na  \cdot \e {\bf E} = \rho$, instead of
$\na  \cdot {\bf E} = \epsilon ^{ - 1} \rho$. The $\e$ cannot move out from the inside of divergence.\\

Except for the fact that $\e$ is matrix, Eq.(3) resembles an isotropic case. It is and equation for $\bf E$, but it is important to note that the $\e$ in the curl of Eq.(4) cannot come to the front of the curl operator, as a constant number does. If the matrix satisfying $A\cd \na \times \bf{E} = \na \times \e \cd \bf{E}$ is found, Maxwell's equation becomes applicable. However there is unfortunately no matrix like $A$.  Therefore, $\e \bf E$ in the curl term serves as an independent variable in a practical sense. Thus, $\bf E$ and $\bf H$ are coupled in the equation, which is not solvable by itself. Nevertheless Eq.(3) can be analytically solved using
tensor Green's functions. The tensor Green's function is explained in the
appendix to check if it satisfies the Green's function condition and
whether or not it solves the problem.\\

After obtaining the solution E, the result is substituted into Eq.(4). A Solution for $\bf H$ then becomes possible with the Green's function technique as well. In this case, the term $(\na \times \e_r {\bf E})$ on the right side of Eq.(4) becomes a part of the source terms, as $\bf{E}$ has already been solved from Eq.(3).\\

To solve the equations, it is convenient to assume that the time dependency is harmonic, or $e^{i\o t}$, and to use the following notations for the source terms of Eqs.(3) and (4):

\by {\bf U}({\bf r}) &=&  - i\o \mu _0 \mu _r {\bf J},\\
{\bf V}({\bf r}) &=&  - \na  \times {\bf J} - i\o \e_0
\na  \times (\e_r {\bf E}). \ey

\ni The corresponding Green's functions satisfies the following conditions:

\by &&\na \times \na \times GE^{(1)}({\bf r},{\bf r'}) + k_0^2
\; GE^{(1)}({\bf r},{\bf r'})\nn\\
&&\;\;\; \;\;\;\;\; = \delta ({\bf r} - {\bf r'}), \\
&&\na ^2 GH^{(1)}({\bf r},{\bf r'}) = \delta ({\bf r} - {\bf r'}). \ey

\ni $GE$ and $GH$ are the Green's functions for an electric and a magnetic field, respectively. The superscript (1) in $GE$ and $GH$ serves to distinguish the anisotropy that occurs. 1 denotes this for $\e$, 2 for $\m$, and 3 denotes this for both  $\e$ and $\m$. The Fourier transform is effective for an unbound case:

\bz GE^{(1)}({\bf r},{\bf r'}) = \iiint {gE^{(1)}({\bf k})\; e^{-i\bf
k  \cd ({\bf r}-{\bf r'})} d^3 {\bf k}} .\ez

Inserting the Fourier-transformed function in Eq.(9) into Eq.(7), we obtain the algebraic equations 

\bz gE^{(1)}({\bf k}) = \frac{1}{{( (k^2{\bf I} - {\bf k} \otimes {\bf k})
- k_0^2 \mu _r \epsilon _r )}}. \ez

\ni Here, $ \otimes$ refers to the direct product, and I is an identity matrix. The above equation for $gE^{(1)}(\bf{k})$ is the main step involved in the solution. Cottis et al. calculated this Green's functions in cylindrical coordinates \cite{Cottis}. The $GE$ for the Green's functions  could be calculated by inserting Eq.(10) into Eq.(9).\\

The form of Eq.(8) is identical to the Poisson equation in electrostatics. The following equation is, thus, applicable:

\bz GH^{(1)}({\bf r},{\bf r'}) =  - \frac{I}{{4\pi \left| {{\bf r} - {\bf
r'}} \right|}}. \ez

\ni The solutions can then be written as follows according to the usual
Green's functions method:

\by {\bf E}({\bf r}) &=& \iiint{GE^{(1)}({\bf r},{\bf r'}){\bf
U}({\bf r'})d^3 {\bf r'}}, \\
{\bf H}({\bf r}) &=& \iiint {GH^{(1)}({\bf r},{\bf r'}){\bf V}({\bf r'})d^3
  {\bf r'}}.  \ey

Looking at Eq.(3), $(-\na ^2 {\bf E})$  is concealed in $(\na \times \na \times {\bf E})$, and the equation is a type of wave equation. As $\e_r$ is a matrix, $E_1$, $E_2$ and $E_3$ are coupled to each other. Therefore, this equation is fundamentally simultaneous PDE. If $\e_r$ is diagonalized,  $E_i\;'$s becomes decoupled after replacing $(\na \times \na \times {\bf
E})$ by $\na (\na \cd {\bf E})-\na ^2 {\bf E}$ . If the
eigenvalues are $\l_1,\l_2 \; and\; \l_3$, the wave vector changes from
$k_i$ to $k_i \sqrt {\l_i}$  on the principal axis. Therefore,
we can compute its refractive index or change of the wave velocity along the corresponding axis.

\section{Anisotropy From permeability }
When anisotropy comes only from $\m_r$, $\m_r$ becomes a matrix, and
$\e_r$ remains a scalar number. The basic equations are as follows:

\by &&\na  \times \na  \times {\bf E} =  - i\o \mu _0
(\na \times (\mu _r {\bf H})), \\
 &&\na ^2 {\bf H} + k_0^2
\epsilon _r \mu _r {\bf H} =  - \na \times {\bf J}. \ey

\ni As in the previous section, $\na  \times (\mu _r
{\bf H})$ in Eq.(14) makes the problem complex. This equation
cannot be solved directly. However, Eq.(15) is merely
Helmholtz-type equations with the source term, $ - \na  \times
{\bf J}$. The calculation is carried out by using a Fourier transform,
as in the preceding section. The Green's functions for Eqs.(14) and (15) are as follows:

\by &&\na  \times \na  \times GE^{(2)}({\bf r},{\bf r'}) = \delta
({\bf r} - {\bf r'}), \\
&&\na ^2 GH^{(2)}({\bf r},{\bf r'}) + k_0^2
\epsilon _r \mu _r GH^{(2)}({\bf r},{\bf r'}) \nn\\
&& \;\;\;\;\;\;\;\;\; = \delta ({\bf r} - {\bf
r'}). \ey

The introduction of the Green's functions $GE$ and $GH$ proceeds identically as it did before. Eq.(17) is a general Helmholtz-type equation of the type studied by many authors \cite{helm}. The Fourier-transformed Green's function $gE^{(2)}(k)$ and $gH^{(2)}(k)$ are defined as follows:

\by GE^{(2)}({\bf r},{\bf r'}) = \iiint {gE^{(2)}({\bf k})e^{-i \bf{k}\cd({\bf r}-{\bf r'})} d^3 {\bf k}}\; , \\
    GH^{(2)}({\bf r},{\bf r'}) = \iiint {gH^{(2)}({\bf k})e^{-i{\bf k}\cd({\bf r}-{\bf r'})} d^3 {\bf k}}\; . \ey

Hence, the algebraic form of $gE^{(2)}(k)$, $gH^{(2)}(k)$ is obtained through insertions into Eq.(16) and (17):

\by &&gE^{(2)}({\bf k}) = \frac{1}{{ (k^2{\bf I} - {\bf k} \otimes {\bf
k})}}\; , \\
&&gH^{(2)}({\bf k},{\bf r'}) = \frac{1}{{(\epsilon _r \mu
_r k_0 ^2 - {\bf I}k_{}^2 )}}\; . \ey

\ni where the denominator is also a matrix showing a tensor property. At this point, the solution for $\bf H$ is possible using $GH^{(2)}({\bf r},{\bf r'})$:

\bz {\bf H}({\bf r}) = \iiint {GH^{(2)}({\bf r},{\bf r'})\; {\bf
V}({\bf r'})d^3 {\bf r'}}. \ez

The source term in the above equation is ${\bf V}({\bf r}) =  - \na  \times
{\bf J}$. The next step is to find the solution. As the function for $\bf H$, has already been solved, the source term of Eq.(14) becomes ${\bf U}({\bf r}) =  -
i\o \mu _0 (\na  \times (\mu _r {\bf H}))$. The solution $\bf E$ is then

\bz {\bf E}({\bf r}) = \iiint {GE^{(2)}({\bf r},{\bf r'})\;{\bf
U}({\bf r'})d^3 {\bf r'}}. \ez

Now, the fields ${\bf E}$ and ${\bf H}$ are solved when the anisotropic characteristic comes from either $\e_r$ or $\m_r$.  Problems  were noted in dealing with the curl term including an anisotropic factor. However, this difficulty is circumvented by solving the other equation first and by placing the result into the curl term, which causes problem. The solutions can
be obtained, but the equations loose their original wave shapes. This does not mean they are not a wave, but the analytical wave nature is not directly observable.

\section{Anisotropy From permittivity and permeability }

When anisotropy occurs due to  both $\e$ and $\m$, the wave equations
become more complicated to solve. This difficulty arises in the same way.
The terms $\e_r \cd \bf{E}$ and $\m_r \cd \bf{H}$ are merely a linear combination of the original $\bf{E}$ and $\bf{H}$, but when contained inside the  curl term $(\na \times )$ , they act as independent variables.  For example, when there is a term $(\na  \times {\bf E})$, we can replace it by $\left( { - \frac{{\pa {\bf B}}}{{\pa t}}} \right)
$ using Maxwell's equations. However, this replacement cannot be applied to
the term $ (\na  \times \epsilon _r {\bf E})$ . It is
practically another unknown to extent that $\bf E$ is solved. The basic wave equations take the following forms:

\by &&\na  \times (\na  \times {\bf E}) =  - i\o \mu _0
\na \times (\mu _r {\bf H}), \\
&&\na ^2 {\bf H} =  - \na
\times {\bf J} - i\o \epsilon _0 \na  \times (\epsilon
_r {\bf E}). \ey

The goal here is clearly to find ${\bf E}$ and ${\bf H}$, but there are two more terms that invoke problems, $(\na  \times \mu _r {\bf H})$ and
$(\na  \times \epsilon _r {\bf E})$. Therefore, it is impossible to obtain the analytical solution simultaneously from the wave equations. A better approach is to go back to the original Maxwell's equations and derive new equations instead of relying on the original wave equations:

\by &&\na \times {\bf E} = -\m_0 \frac{\pa (\m_r {\bf H})}
{\pa t}, \\ &&\na \times {\bf H} = J + \e_0 \frac{\pa (\e_r {\bf E})}{\pa t}.\ey

The magnetic field ${\bf H}$ is written from Eq.(26) as
\bz {\bf H} = \int \frac{-1}{\m_0} \; \m_r^{-1}\cd (\na \times {\bf E})\; dt.\ez
Inserting the above equations into Eq.(27) and differentiating with respect to time gives
\bz \na\times (\m_r^{-1} \cd (\na\times {\bf E})) +\frac{1}{c^2} \; \e_r \frac{\pa^2 {\bf E}}{\pa t^2} = - \m_0 \fr{\pa {\bf J}}{\pa t}\; .\ez

\ni This equation restores the original wave equations when the media is isotropic. Moreover, it is an equation for $\bf E$ that is decoupled from $\bf H$. In fact, this equation holds in any case, regardless of the existence of anisotropy. If the time dependency in ${\bf E}$ and ${\bf J}$ are assumed to be harmonic or $e^{i\o t}$, the above equations read.
\bz \na\times (\m_r^{-1} \cd (\na\times {\bf E})) - k_0^2 \; \e_r {\bf E} = - i \o \m_0 {\bf J}\ez

As the equation is linear, the tensor Green's function 
\bz \na\times (\m_r^{-1} \cd (\na\times GE^{(3)}({\bf r},{\bf r'}))) - k_0^2 \; \e_r GE^{(3)}({\bf r},{\bf r'})= \d({\bf r} - {\bf r'})\ez
is also possible. Assuming that $gE^{(3)}({\bf r},{\bf r'})$ is Fourier-transformed  in the same way to the Eq.(18) in the last section, the spatial Green's function
\bz GE^{(3)}({\bf r},{\bf r'}) = \iiint {gE^{(3)}({\bf k})e^{-i \bf{k}\cd({\bf r}-{\bf r'})} d^3 {\bf k}}. \ez

Eq.(31) provides the algebraic relationship for $gE^{(3)}$:
\bz -{\bf k}\times \m_r^{-1} \times ({\bf k} \times gE^{(3)}) -k_0^2 \e_r gE^{(3)}  = 1. \ez

\ni By applying the following matrix $\tilde k$ representing the wave vector $\bf k$ \cite{abdul},
\bz \tilde k = \left( {\begin{array}{*{20}c}
   0 & { - k_z } & {k_y }  \\
   {k_z } & 0 & { - k_x }  \\
   { - k_y } & {k_x } & 0  \\
\end{array}} \right),
\ez

Eq.(33) above becomes
\bz gE^{(3)}({\bf k}) = \frac{-1}{{\tilde k}\; \m_r^{-1}{\tilde k} +k_0^2 \; \e_r}\; .\ez

Here, $\tilde k$ is defined in rectangular coordinates. This implies that $\tilde k$ is coordinates dependent. Now, $gE^{(3)}({\bf k})$ is obtained, and the calculation of $GE^{(3)}$ and ${\bf E}({\bf r})$ are straightforward.

\by &&GE^{(3)}({\bf r},{\bf r'}) = \iiint {gE^{(3)}({\bf k})\; e^{-i\bf
k  \cd ({\bf r}-{\bf r'})} d^3 {\bf k}}\; , \\
&&{\bf E}({\bf r}) = \iiint{GE^{(3)}({\bf r},{\bf r'})\;{\bf
U}({\bf r'})d^3 {\bf r'}}\; , \ey

\ni where ${\bf U}({\bf r})$ is the right term of Eq.(30) as a source.
The solution of $\bf H({\bf r})$ is obtained directly from Eq.(28) by inserting $\bf E$. In this section, a classical wave equation itself was not considered. Instead, a derivation of new solvable wave equations from Maxwell's equations was presented.

\section{Discussion and conclusion}
When anisotropy occurs due to either permittivity or permeability, the wave equation is solved by using Green's functions. However, in an arbitrary case in which anisotropy occurs in both $\e_r$ and $\m_r$, the equation is not substantially solvable in an analytic sense. The usual means is to resort to a numerical calculation.\\

However, there is another method that gradually approaches solutions by iteration. This does not need to be a numerical calculation. It can be performed analytically. Either $\e_r$ or $\m_r$ becomes a scalar number; then, the whole system can be solved as before. Therefore, the solution of the equations becomes feasible by replacing $\e_r$ with a scalar number. The solution obtained in this way is the first trial solution. The equations are then solved again by inserting the first solutions into the curl term, but from this time, the original matrix value for $\e_r$ is used instead of the first number that was chosen at the beginning. An improved second solution for $\bf E$ and $\bf H$ is then calculated. If this iteration continues, the results approach some converging functions, which are supposed to be solutions to the equations. This procedure can be carried out until difference between the $(n-1)^{th}$ solution and the $n^{th}$ solution becomes less than a prescribed level.This process can be done by using a symbolic calculation \cite{sym}. However, it is merely an approximate method, regardless of whether it is a numeric or a symbolic calculation.\\

Instead of solving the original equations, we derived new wave equations that were especially useful for an anisotropic problem via a slight manipulation of Maxwell's equations. The new equation is identical to the original wave equation when there is no anisotropic characteristic. The fundamental advantage of this new equation is the fact that $\bf E$ and $\bf H$ are decoupled, unlike in the original equations, making it possible to find analytical propagators.\\

In an actual calculation, there is one factor to consider; i.e. Eq.(35) becomes singular in the isotropic case. The determinant of the denominator becomes zero in this special case when $\e_r =\m_r = scalar\; number$. In such an isotropic case, the solution can be obtained in an ordinary manner. \\

Eq.(29) was derived for an electric field. An equation for the magnetic field can also be obtained in a similar way. The result is as follows:
\bz \na\times \e_r^{-1} (\na \times {\bf H}) +\fr{1}{c^2} \m_r \;
\frac{\pa^2 {\bf H}}{\pa t^2}
= \na\times (\e_r^{-1} {\bf J}). \ez
Solving this equation is identical to the procedure for $\bf E$ given in the last section.\\

The new wave equations can be justified in several ways. The best approach is to compare the experimental values with the calculated result for the new equations. However, many aspects can be examined analytically by comparing the results of the two systems of the equations. The first comparison is that the two equation sets are identical in an isotropic case. these can be seen immediately by inspection, because there is no matrix. For anisotropic cases, the Fourier-transformed Green's functions can be checked whether they are identical or not. Cottis, Vazouras and Spyrou calculated the Fourier-transformed dyadic Green's functions in a dielectric anisotropy, and their result is the same as Eq.(10). The result of the new equation for an identical case is given below according to Eq.(35):
\bz gE^{(3)}({\bf k}) = \frac{-1}{{\tilde k}\; {\tilde k} +k_0^2 \; \e_r}\; .\ez
Given that there is no anisotropy in $\m$, the identity matrix is substituted in place of $\m_r^{-1}$. The same result is obtained using the relationship $\tilde k \cd \tilde k \; = \; {\bf k}\otimes {\bf k}-k^2 I$.\\

The other comparison is for the case of magnetic anisotropy. In this case $\e_r = I$. The result of the Fourier-transformed Green's functions of the original equation is given in Eq.(21). It can be compared with the result of Eq.(38), which gives 

\by gH^{(3)} &&= \fr{-1}{\tilde k \; \e_r^-1 \; \tilde k + k_0^2 \; \m_r} \nn\\
&&= \fr{-1}{\tilde k \; I \; \tilde k + k_0^2 \;\m_r} \nn\\ 
&&=  \fr{-1}{{\bf k} \otimes {\bf k} - k^2 I + k_0^2 \;\m_r}\; . \ey
However, the ${\bf k} \otimes {\bf k}$ term becomes zero when it is applied to a magnetic field. Hence, the result of the original equation is identical to that of the new equations.\\

For more verification, the plane wave ${\bf E} = {\bf E}_0 \; exp[i({\bf k\cd r}-\o t)]$ can be considered in an anisotropic medium. When the relative permittivity tensor is given in terms of refractive indices as 
\bz \e_r \; =\;
\left( {\begin{array}{*{20}c}
   n_x^2 & 0 & 0  \\
   0 & n_y^2 & 0  \\
   0 & 0 & n_z^2  \\
\end{array}} \right), \ez
the refractive indices then satisfy the following well-known condition:
\bz 
\left| {\begin{array}{*{20}c}
   n_x^2 -n^2 (\hat{k}_y^2 \hat{k}_z^2 )& n^2 \hat{k}_x \hat{k}_y & n^2 \hat{k}_x \hat{k}_z  \\
   n^2 \hat{k}_x \hat{k}_y & n_y^2 -n^2 (\hat{k}_x^2 +\hat{k}_z^2 ) & n^2 \hat{k}_y \hat{k}_z  \\
   n^2 \hat{k}_x \hat{k}_y & n^2 \hat{k}_y \hat{k}_z & n_z^2 -n^2 (\hat{k}_x^2 \hat{k}_y^2 )  \\
   \end{array}} \right| 
    \;=\; 0 , \ez
where the conventional notations ${\bf n}=\fr{c}{\o} {\bf k}$ and $\hat{k}_i =\fr{k_i}{\|{\bf k}\|}$ are used.
Exactly the same result is also obtained from the new wave equation, Eq.(29). This is actually a natural consequences, because they are based on the same mathematical ground.\\

As for advantages of the new equations, it is clear that original equations do not allow an analytical approach in the general case whereas the new equations give analytical propagators (Fourier-transformed Green's functions). The point is whether or not they are decoupled. This is a considerable difference between the two systems, although the new equations need numerical calculations, for example integration to get $GE({\bf r},{\bf r'})$, $GH({\bf r},{\bf r'})$ or the final answers $\bf E$ and $\bf H$. The existence of propagators is supposed to bring a non-negligible amount of code-saving effect.\\

As a result, an equation set Eqs.(29) and (38), is obtained and describes wave propagation in anisotropic media. This shows that the two systems are commutable and equivalent. The new wave equation is expected to be useful in analytical research of anisotropic properties beyond what a numerical approach can address.

\appendix
\section{Proof of the Tensor Green's function}

As Eq.(3) is linear, it is natural to consider
Green's functions. However, the equation is a multi-linear simultaneous
PDE, and can verify if the use of Green's functions is proper. If Green's functions work, the following type of solutions can be assumed as valid:

\bz {\bf E}_i  = \iiint {GE_{ij} {\bf U}_j d{\bf r'}}\; . \ez

The functions $GE_{ij}$ are devised to play the roles of influence
functions to generate the fields due to $\rho$ and $\bf J$.  At this
stage, it is unknown whether or not they are Green's functions. The proof should be performed as to whether they satisfy the conditions of Green's functions before adapt ion of Green's functions. Eq.(3) can be written in component form as 

\bz \epsilon _{ikq} \epsilon _{klm} \frac{{\pa ^2 E_1
}}{{\pa x_m \pa x_q }} + \frac{1}{{c^2 }}\mu _r
(\epsilon _r )_{ij} E_j  =  - i\o \mu _0 \mu _r J_i. \ez

By inserting Eq.(A1) into Eq.(A2), one can calculate the next equation:

\bz \int {\{ \epsilon _{ikq} \epsilon _{klm}
\frac{{\pa ^2 GE_{1s} }}{{\pa x_m \pa x_q }} +
\frac{1}{{c^2 }}\mu _r (\epsilon _r )_{ij} GE_{js} \} U_s d{\bf
r}' = U_i^{} }\; ,  \ez

\ni where $U_i  =  - i\o \mu _0 \mu _r J_i $ for the source term. Inside the bracket is $\na  \times \na  \times GE - k_0^2 \mu
_r \epsilon _r GE$, and to maintain the equality, as expected, the equation
arrives at the definition of the Green's functions:

\bz \na  \times \na  \times GE - k_0^2 \mu _r \epsilon _r
\frac{\pa ^2 GE}{\pa t^2 } = I\delta ({\bf r} - {\bf
r'})\; . \ez

\ni $GE$ and I are 3-by-3 matrices and the above equation is simply a definition of the Green's function.

The other tensor Green's functions used in this article all
have similar forms. Moreover, it is easy to prove that they meet the definition of the Green's function. The equation in Eq.(A4) is essentially nine equations for $GE_{ij}$ that are coupled to each other. However, the equations become decoupled and much simpler to solve through diagonalization of $\e_r$.

\end{document}